\documentclass[aps,showpacs]{revtex4}

\usepackage{graphicx}
\usepackage{dcolumn}
\usepackage{bm}
\usepackage{epsfig}
\newcommand{\be}{\begin{eqnarray}}
\newcommand{\ee}{\end{eqnarray}}

\def\d{\mbox d}

\begin{document}


\title{Ferromagnetic behaviour in the strongly interacting two-component Bose gas}

\author{Xi-Wen Guan${}^{1}$, Murray T. Batchelor${}^{1,2}$ and Minoru Takahashi${}^{3}$}
\affiliation{${}^{1}$Department of Theoretical Physics, Research School of Physical Sciences and Engineering,\\
Australian National University, Canberra ACT 0200,  Australia\\
${}^{2}$Mathematical Sciences Institute, Australian National University, Canberra ACT 0200,  Australia\\
${}^{3}$Department of Physics, Toho University, Miyama 2-2-1, Funabashi 
274-8510, Japan}
\date{\today}

\begin{abstract}
\noindent
We investigate the low temperature behaviour of the integrable 1D 
two-component spinor Bose gas using the thermodynamic Bethe ansatz. 
We find that for strong coupling the characteristics of the thermodynamics at low
temperatures are quantitatively affected by the spin ferromagnetic states, 
which are described by an effective ferromagnetic Heisenberg chain. 
The free energy, specific heat, susceptibility and local pair correlation function are calculated 
for various physical regimes in terms of temperature and interaction strength. 
These thermodynamic properties reveal spin effects which are significantly different 
than those of the spinless Bose gas. 
The zero-field susceptibility for finite strong repulsion exceeds that of a free spin paramagnet. 
The critical exponents of  the specific heat $c_v \sim T^{1/2}$  and the susceptibility 
$\chi \sim T^{-2}$ are indicative of the ferromagnetic signature of the two-component spinor Bose gas. 
Our analytic results are consistent with general arguments by Eisenberg and Lieb for polarized 
spinor bosons.

\end{abstract}



\maketitle 

\section{Introduction}

Experiments with ultracold quantum gases are opening up exciting new possibilities for testing 
and exploring quantum effects in many-body systems (for recent reviews, see Refs.~\cite{Lewenstein,Grimm,Stringari}). 
These include experiments on effectively one-dimensional (1D) quantum Bose gases of $^{87}$Rb 
atoms in which the interaction strength between atoms is tunable \cite{E-TG1,E-TG2,E-TG3,E-TG4,E-TG5}.
The experiments provide a striking example of realizing an integrable quantum many-body problem.
They demonstrate the explicit fermionization of bosons and provide a direct test of 
theoretical results obtained for the integrable 1D interacting (spinless) Bose gas \cite{LL,JMB}.  
Another frontier of activity involves spinor Bose gases of alkali atoms in which 
hyperfine states comprise the pseudospins \cite{Exp-SB1,Exp-SB2,Exp-SB3}.  
In these systems quantum collisional effects can produce
spatio-temporal spin oscillations (spin waves) \cite{spinor,YangK,Ohmi,Ashhab}. 
The observation of collective dynamics of spin waves and spin-state
segregation in trapped spinor Bose gases has stimulated
a wide range of interest in studying magnetism, topological spin
defects and novel quantum phase transitions in spinor Bose gases \cite{Rev,Ueda}.

Two-component spinor Bose gases have been experimentally created in a
magnetic trap by rotating two hyperfine states so that the two
atomic hyperfine states make up a pseudo-spin doublet \cite{Spinor-2},
e.g., the $|F=2,m_F=-1\rangle$ and $|F=1,m_F=1\rangle$ hyperfine states of ${}^{87}$Rb.
In general, spin-independent s-wave scattering dominates interactions in alkali atomic gases. 
In 1D, the two-component Bose gas with spin-independent s-wave scattering can be
exactly solved, like the spinless model, by means of the Bethe ansatz
\cite{Sutherland,Li}. 
In contrast to Fermi gases, ferromagnetic order emerges in spinor
Bose gases as long as the interaction is fully spin independent \cite{Eisenberg-Lieb,YangK}.  
The low-energy excitations of the model split into collective excitations carrying charge 
and collective excitations carrying spin. 
The charge excitations are phonons whereas the spin excitations have quadratic dispersion 
connected to spin wave excitations \cite{Fuchs,BBGO}. 
Spin dynamics in the 1D ferromagnetic Bose gas have been studied recently \cite{Zvonarev}.
Girardeau's Fermi-Bose mapping has been used to study the 1D spinor
Bose gases \cite{Deuretzbacher}.
In general the two-component interacting Bose gas provides a tunable testing ground for observing 
the phenomenon of spin-charge separation \cite{S-C}.

Quantum gases with multi-spin states are expected to exhibit even richer quantum effects
than their single component counterparts \cite{Lewenstein,Grimm,Stringari}.
Universal features appearing in the low temperature behaviour of 
strongly interacting spinor Bose gases should differ significantly from
those of spinless Bose gases and the antiferromagnetic
behaviour of Fermi gases due to their fundamentally different statistical signatures.  
One way to calculate the thermodynamics of integrable many-body systems
is via the thermodynamic Bethe ansatz (TBA)  \cite{Korepin,Takahashi,Hubbardbook,Kondo,Schlot}, 
introduced by Yang and Yang \cite{Yang-Yang} for the 1D Bose gas.
However, it is a challenging problem to derive exact TBA results for the thermodynamics
of 1D quantum many-body systems. 
Our aim here is to obtain universal characteristics of ferromagnetic behaviour for 
the 1D two-component strongly interacting Bose gas of ultracold atoms
via the TBA method.  
We will see that the ferromagnetic phase associated with the spin degrees of
freedom may separate from the gas phase in the strongly repulsive regime
due to spin-charge separation.  
The low temperature behaviour is dominated by the spin ferromagnetic states, 
which are described by an effective ferromagnetic Heisenberg chain. 
In this way we make contact with the known results for the thermodynamics 
of the ferromagnetic Heisenberg chain \cite{Takahashi2,Schlot2} to derive analytic expressions for the free energy,
specific heat, susceptibility and local pair correlation function for the
strongly interacting two-component Bose gas in terms of temperature and interaction strength.  
These thermodynamic properties reveal some novel spin effects.  
Our explicit results are consistent with general arguments by Eisenberg and Lieb  \cite{Eisenberg-Lieb}
for polarized spinor bosons.

This paper is set out as follows. In section \ref{sec:model} we present
the Bethe ansatz solution of the 1D two-component interacting Bose gas. 
The ground state properties are also calculated. 
In section \ref{sec:TBA} we introduce the TBA for the spinor Bose gas in order to
study the thermodynamics at low temperatures, including the analysis of spin charge separation.  
The ferromagnetic ground state is studied by means of the solution of the TBA equations in section 
\ref{sec:GS-LL}. We discuss low temperature ferromagnetic behaviour for
the 1D strongly interacting Bose gas of atoms in section \ref{sec:LT}. The
local pair correlation function is studied at low temperatures in section \ref{sec:pair}.
Section \ref{sec:concl} is devoted to a brief summary and concluding remarks.

\section{The two-component spinor Bose gas}
\label{sec:model}

The Hamiltonian describing a $\delta$-function interacting gas of $N$ bosons of mass
$m$ constrained by periodic boundary conditions to a line of length $L$ 
with internal degrees of freedom is 
\begin{equation}
{\cal H}=-\frac{\hbar ^2}{2m}\sum_{i = 1}^{N}\frac{\partial ^2}{\partial x_i^2}
+\,g_{\rm 1D} \sum_{1\leq i<j\leq N} \delta (x_i-x_j).
\label{Ham-1}
\end{equation}
For the ultracold atomic gases  \cite{Olshanii}, the coupling constant $g_{\rm 1D}$ can be written 
in terms of the scattering strength $c={2}/{a_{\rm 1D}}$ as $g_{\rm 1D} ={\hbar ^2 c}/{m}$. 
The effective 1D scattering length $a_{\rm 1D}$ can be related to the 3D scattering length for 
bosons or fermions confined in a $1$D geometry. 
The dimensionless coupling constant $\gamma=c/n={mg_{\rm 1D}}/{(\hbar^2n})$ 
is convenient for physical analysis. Here $n={N}/{L}$ is the linear density. 
We take $2m=\hbar=1$ for simplicity in the following equations.
However, we reinstate them where appropriate in discussing the thermodynamics of the model. 
The wavefunctions of Hamiltonian (\ref{Ham-1}) for the spinor Bose gas are symmetric under 
exchange of spatial and internal spin coordinates between two particles. 
We shall see that this statistical signature triggers rather novel ferromagnetic behaviour in the 
degenerate quantum spinor Bose gas.  
The interaction is attractive for $g_{\rm 1D}<0$ and repulsive for $g_{\rm 1D}>0$. 
However,  one should note that there is no thermodynamic limit for the attractive case in Bose gases.

For $M$ spin-down bosons, the  Bethe ansatz equations (BAE) for the two-component Bose gas 
are of the form \cite{Sutherland,Li}
\begin{eqnarray}
\exp(\mathrm{i}k_jL)&=&-\prod^N_{\ell = 1} 
\frac{k_j-k_\ell+\mathrm{i}\, c}{k_j-k_\ell-\mathrm{i}\, c}\prod^M_{\alpha = 1}
\frac{k_j-\lambda_\alpha-\frac12 \mathrm{i} c}{k_j-\lambda_\alpha+\frac12 \mathrm{i} c}, \qquad j = 1, \ldots, N
\nonumber\\
\prod^N_{\ell = 1}
\frac{\lambda_{\alpha}-k_{\ell}-\frac12 \mathrm{i} c}{\lambda_{\alpha}-k_{\ell}+\frac12 \mathrm{i} c}
 &=& - {\prod^M_{ \beta = 1} }
\frac{\lambda_{\alpha}-\lambda_{\beta} -\mathrm{i}\, c}{\lambda_{\alpha}-\lambda_{\beta} +\mathrm{i}\, c}, 
\qquad \alpha = 1,\ldots, M
\label{BE}
\end{eqnarray}
in terms of which the energy eigenspectrum is given by $E=\sum_{j=1}^N k_j^2$.
In the thermodynamic limit, i.e., $N,L\to \infty$ with $N/L$ finite, these equations can be
written as coupled integral equations in terms of the particle and hole root densities
$\rho(k)$ and $\rho^h(k)$ ($\sigma(\lambda)$ and $\sigma^h(\lambda)$) for the charge (spin) degrees
of freedom, respectively.
These  are
\begin{eqnarray}
\rho(k)+\rho^h(k)&=&\frac{1}{2\pi}+\frac{1}{2\pi}\int_{-Q}^{Q}\frac{2c
  \rho(k')}{c^2+(k-k')^2}dk'
-\frac{1}{2\pi} \int_{-B}^{B}\frac{c\sigma(\lambda)}{c^2/4+(k-\lambda)^2}d\lambda\nonumber\\
\sigma(\lambda)+\sigma^h(\lambda)&=&\frac{1}{2\pi}\int_{-Q}^{Q}\frac{c \rho(k)}{c^2/4+(\lambda-k)^2}dk 
-\frac{1}{2\pi} \int_{-B}^{B}\frac{2c\sigma(\lambda)}{c^2+(\lambda-\lambda')^2}d\lambda'.\label{BA-B}
\end{eqnarray}
The integration limits $Q$ and $B$ are determined by $N/L=\int_{-Q}^Q\rho(k)dk$ and
$M/L=\int_{-B}^{B}\sigma(\lambda)d\lambda$. 
At zero temperature, the ground state corresponds to the configuration $\sigma(\lambda)=\rho^h(k)=0$ 
leading to a ferromagnetic ground state. 
For the ground state there are therefore no holes in the charge degrees of freedom and no 
quasiparticles in the spin degrees of freedom. 
However, as the temperature increases spin strings become 
involved in the thermal equilibrium states.  
We shall investigate this ground state configuration via analysis of the string solutions to the TBA.

\section{The thermodynamic Bethe ansatz}
\label{sec:TBA}

For finite temperatures each of the $N$ quasimomenta $k_i$ are real due to the repulsive interaction.
However, the spin quasimomenta form complex strings of the form \cite{Takahashi3}
$\lambda_{\alpha,j}^n=\Lambda^n_{\alpha}+\mathrm{i}(n+1-2j)c/2$ for $j=1,\ldots ,n$. 
Here the number of strings $\alpha =1,\ldots, N_n$. 
$\Lambda^n_{\alpha}$ on the real axis denotes the position of the centre of a length-$n$ string. 
The number of $n$-strings $N_n$ satisfies the relation $M=\sum_nnN_n$. 
It is assumed that the distribution of Bethe roots along the real axis is dense enough to pass to the 
continuum limit. 
After performing a standard calculation with the string solutions and 
introducing the  convolution integral 
$(f*g)(\lambda) = \int_{-\infty}^\infty f(\lambda-\lambda') g(\lambda') d\lambda'$, 
the BAE (\ref{BE}) become
\begin{eqnarray}
\rho(k)+\rho^h(k)&=&\frac{1}{2\pi}+\frac{1}{2\pi}\int_{-\infty}^{\infty}\frac{2c\rho(k')dk'}{c^2+(k-k')^2}
-\sum_{n=1}^{\infty}a_n*\sigma_n(k)\nonumber\\
\sigma_n(\lambda)+\sigma^h_n(\lambda)&=&a_n*\rho(\lambda)-\sum_{m=1}^{\infty}T_{nm}*\sigma_m(\lambda)
\label{TBA-BE}
\end{eqnarray}
where
\begin{eqnarray}
T_{nm}(\lambda)=\left\{ \begin{array}{l}
  a_{|n-m|}(\lambda)+2a_{|n-m|+2}(\lambda)+\ldots 
 +2a_{n+m-2}(\lambda)+a_{n+m}(\lambda),\,\,{\rm for }\, \,n\ne m,\\
 2a_2(\lambda)+2a_{4}(\lambda)+\ldots
  +2a_{2n-2}(\lambda)+a_{2n}(\lambda),\,\,{\rm for } \,\, n=
  m,\end{array}\right.
\end{eqnarray}
and
\begin{equation}
a_n(\lambda)=\frac{1}{2\pi}\frac{nc}{(nc/2)^2+\lambda ^2}.\label{a-n}
\end{equation}

The equilibrium states at finite temperature  $T$ are described by the
equilibrium particle and hole densities $\rho(k)$ and $\rho^h(k)$ of the 
charge degrees of freedom and the equilibrium string densities
$\sigma_n(\lambda)$ and $\sigma_n^h(\lambda)$ of the spin degrees of freedom.
Here $n=1,2,\ldots, \infty$.
The partition function $Z= {\rm tr}(\mathrm{e}^{-{\cal{H}}/T})$ is defined by
\begin{eqnarray}
Z&=&\sum_{\rho,\rho^h,\sigma_n,\sigma_n^h
}W(\rho,\rho^h,\sigma_n,\sigma_n^h)\mathrm{e}^{-E(\rho,\rho^h,\sigma_n,\sigma_n^h)/T}
\end{eqnarray}
where the densities satisfy (\ref{TBA-BE}) with $W(\rho,\rho^h,\sigma_n,\sigma_n^h)$ the
number of states corresponding to the given densities. 
Introducing the combinatorial entropy $S=\ln W(\rho,\rho^h,\sigma_n,\sigma^h_n)$,
the grand partition function is $Z=\mathrm{e}^{-G/T}$, where the Gibbs
free energy $G=E-\mu N-H(N_{\uparrow}-N_{\downarrow})/2-TS$. 
Here $\mu$ is the chemical potential and $N_{\uparrow}$ ($N_{\downarrow}$)
denotes the number of the particles with up (down) spin.
Recall that $N_{\downarrow}=M$.
The energy per unit length is defined by
\begin{eqnarray}
{E}/{L}= \int_{-\infty}^\infty k^2\rho(k)dk-m^zH.
\end{eqnarray}
Here $H$ is the external magnetic field and
$m^z = (N_{\uparrow}-N_{\downarrow})/2$ denotes the atomic magnetic
momentum (where the Bohr magneton $\mu_B$ and the Lande factore are
absorbed into the magnetic field $H$). 
The magnetization per unit length in  the $z$-direction is thus given by
\begin{equation}
m^z=\frac{1}{2}\int_{-\infty}^\infty\rho(k)dk-\sum_nn\int_{-\infty}^\infty \sigma_n(\lambda)\d\lambda.
\end{equation}

Now the equilibrium states are determined by the minimization condition of
the Gibbs free energy \cite{Yang-Yang,Takahashi}, i.e., the condition $\delta(E-\mu n-TS)=0$, 
which gives rise to the set of coupled nonlinear integral equations (the TBA equations)\cite{TBA-Li}
\begin{eqnarray}
\epsilon(k)&=&k^2-\mu-\frac{1}{2} H
-\frac{T}{2\pi}\int_{-\infty}^\infty\frac{2c}{c^2+(k-k')^2}\ln(1+\mathrm{e}^{-{\epsilon(k')}/{T}})
-T\sum_{n=1}^{\infty}a_n(k-\lambda)*\ln(1+\eta^{-1}_n(\lambda))\nonumber\\
\ln \eta_n(\lambda)&
=&\frac{nH}{T}+a_n(\lambda-k)*\ln(1+\mathrm{e}^{-{\epsilon(k')}/{T}})
 +\sum_{n=1}^{\infty}T_{mn}(\lambda-\lambda')*\ln(1+\eta^{-1}_n(\lambda')).\label{TBA}
\end{eqnarray}
Here we have defined the dressed energy $\epsilon(k):=
T\ln(\rho^h(k)/\rho(k))$ with respect to the quasimomentum $k$. 
Similarly $\eta(\lambda):=\sigma^h(\lambda)/\sigma(\lambda)$ with respect
to the spin quasimomentum $\lambda$.  
The dressed energy $\epsilon(k)$ plays the role of excitation energy measured from the Fermi energy
level $\epsilon(k_{\rm F})=0$, where $k_{\rm F}$ is the Fermi momentum.  
The pressure $P(T,H)$ and free energy $F(T,H)$ per unit length in the thermodynamic limit are 
given in terms of the dressed energy by 
\begin{eqnarray}
P(T,H)&=&\frac{T}{2\pi}\int_{-\infty}^\infty dk\ln(1+\mathrm{e}^{-{\epsilon(k)}/{T}})
\label{Pressure}\\
F(T,H)&=&\mu n-\frac{T}{2\pi}\int_{-\infty}^\infty dk \ln(1+\mathrm{e}^{-{\epsilon(k)}/{T}}).
\label{Free-E}
\end{eqnarray}

At low temperatures and zero magnetic field, it is important to note
that $\mathrm{e}^{-\epsilon(k)/T}$ is negligibly small if the energy
$\frac{\hbar^2 k^2}{2m} > 2 \mu$, where $\mu$ is the chemical potential. 
It follows that for strong coupling, i.e., for $\gamma \gg 1$, the TBA equations (\ref{TBA}) 
can be written as 
\begin{eqnarray}
\epsilon(k)&=&k^2-\mu-\frac{1}{2} H -\frac{2c P(T,H)}{c^2+k^2}
-T\sum_{n=1}^{\infty}\int_{-\infty}^{\infty}a_n(\lambda )\,\ln(1+\eta_n^{-1}(\lambda)) \nonumber\\
\ln \eta_n(\lambda)& =&\frac{nH}{T}+\frac{4\pi P(T,H)
  a_n(\lambda)}{cT}
  +\sum_{n=1}^{\infty}T_{mn}(\lambda-\lambda')*\ln(1+\eta^{-1}_n(\lambda')).
  \label{TBA-3}
\end{eqnarray}
In the above equations, we have only kept terms to order $1/c$ for the dressed enery $\epsilon(k)$.  
We have also made the variable changes $\lambda \to c\lambda/2$ and $\eta(\lambda) \to \eta(c\lambda/2)$.  
Thus the function $a_n(\lambda)$ defined in (\ref{a-n}) becomes $a_n(x)=\frac{1}{\pi}\frac{n}{n^2+x^2}$, 
which is Takahashi's notation for the ferromagnetic Heisenberg chain \cite{Takahashi}.  
It will be clearly seen from the TBA equations (\ref{TBA-3}) that the 
low temperature behaviour of the strongly interacting two-component boson model 
is determined by the hard core Bose gas state and ferromagnetic spin wave fluctuations 
described by the ferromagnetic Heisenberg chain with an effective coupling strength $J=2P(T,H)/c>0$.
We shall see through the TBA equations  (\ref{TBA-3}) that the temperature dependent part of this effective 
coupling strength contributes to the free energy at $O(T^3)$. 
At low temperatures $J\to 4E_{\rm F}/3$, where $E_{\rm F}$ is the Fermi energy.

Following the analysis of the TBA equations for the ferromagnetic Heisenberg chain \cite{Takahashi}, 
the TBA equations (\ref{TBA-3}) can be rewritten in terms of the function $\eta_1(\lambda)$ in the form
\begin{equation}
\epsilon(k) \approx k^2-\mu-\frac{2cP(T,H)}{c^2+k^2}+f_{XXX}(T,H)
 \label{TBA-HB}
\end{equation}
where 
\begin{equation}
f_{XXX}(T,H)\approx J\ln 2-T\int_{-\infty}^{\infty}d\lambda s(\lambda)\ln(1+\eta_1(\lambda))\label{TBA-XXX-f}
\end{equation}
is the free energy of a ferromagnetic Heisenberg chain with
coupling $J\approx 2P(T,H)/c$. Here $1/s(\lambda)={4\cosh(\pi \lambda/2)}$.  
The function $\eta_1(\lambda)$ is determined by the TBA equations for the ferromagnetic 
Heisenberg chain \cite{Takahashi}, which are
\begin{eqnarray}
\ln \eta_1(\lambda)& =& \frac{2\pi J}{T}s(\lambda)+s*\ln(1+\eta _2(\lambda))\nonumber\\
\ln \eta_n(\lambda)& =& s*\ln(1+\eta _{n-1}(\lambda))\ln(1+\eta _{n+1}(\lambda)).\label{TBA-XXX}
\end{eqnarray}
For large $n$, the string solutions obey $\lim_{n\to \infty} n^{-1}{\ln \eta_n(\lambda )}={H}/{T}$.

So far we have separated the ferromagnetic spin state from the hard core gas phase via the TBA formalism. 
It is important to note that for the strong coupling regime the 
spin and charge degrees of freedom are coupled via the pressure. 
The spin velocity vanishes whereas the charge velocity tends to the Fermi velocity of noninteracting spinless
fermions as $\gamma \to \infty$. 
In this extreme case, the effective mass takes the
maximum value $m^*=Nm$, meaning that by moving one boson with down
spin, one has to move all the particles with up spins \cite{Fuchs,BBGO}.  
For finitely strong repulsion, a quadratic dispersion of
spin wave excitations above the ferromagnetic ground state can be
obtained from the Bethe ansatz solution (\ref{BE}) \cite{Fuchs,BBGO}.  
In general spin charge separation is a typical phenomenon in interacting many-body systems \cite{S-C}.

Before moving on to discuss the solutions of the TBA (\ref{TBA-XXX}), we proceed to calculate the
pressure (\ref{Pressure}). 
Without loss of generality, we let $\epsilon(k)=\frac{\hbar^2 k^2}{2m}-A(T,H)$, where
$A(T,H)=\mu+{2P(T,H)}/{c}-f_{XXX}(T,H)$.  
Hereafter we restore physical units in the thermodynamic properties. 
Integration by parts gives 
 \begin{eqnarray}
P(T,H)\approx
\frac{1}{\sqrt{\frac{\pi^2\hbar^2}{2m}}}\int_0^{\infty}\frac{\sqrt{\epsilon} \, d\epsilon
}{1+\mathrm{e}^{\frac{\epsilon-A(T,H)}{K_BT}}}.\label{Pressure-2}
\end{eqnarray}
The integral in (\ref{Pressure-2}) can be calculated explicitly using Sommerfeld
expansion \cite{book} to give
\begin{eqnarray}
P(T,H)&\approx&
\frac{1}{\sqrt{\frac{\pi^2\hbar^2}{2m}}} \frac{2}{3} A(T,H)^{\frac{3}{2}}
\left[1+
\frac{\pi^2}{8}\left(\frac{K_BT}{A(T,H)}\right)^2 +\frac{7\pi^4}{640}\left(\frac{K_BT}{A(T,H)}\right)^4+\cdots \right].
\label{Pressure-3}
\end{eqnarray}
Here the function $A(T,H)$ contains $P(T,H)$. 
{}From equation (\ref{Pressure-3}) we can find a relation between the pressure $P(T,H) $ and chemical
potential $\mu$ by iteration.
This provides a starting point to calculate the thermodynamics of the strongly interacting two-component Bose 
gas at low temperatures, including the zero temperature limit.


\section{The ground state and the spinless Bose gas}
\label{sec:GS-LL}

Pure dynamical interaction drives the spinless Bose gas into distinct quantum phases of matter:
from the quasi-Bose-Einstein condensate to the Tonks-Girardeau phase.  
As stated in the introduction, this elegantly simple Bethe ansatz solved 1D quantum many-body 
system \cite{LL,JMB} is testable in experiments on trapped quantum gases of
ultracold atoms \cite{E-TG1,E-TG2,E-TG3,E-TG4}. 
It is natural to expect that spin dynamics in the interacting two-component spinor Bose gases would lead to
significantly different quantum effects than those of the spinless Bose gas. 
At $T\to 0$, it is suitable to use the dressed energy formalism in the TBA equations (\ref{TBA-XXX}), i.e.,
\begin{eqnarray}
\xi_1(\lambda)&=&2\pi Js(\lambda)+s*\xi^+_2(\lambda)\nonumber\\
\xi_n(\lambda)&=&s*\left(\xi^+_{n-1}(\lambda)+\xi^+_{n+1}(\lambda)\right)
 \label{TBA-XXX-T0}
\end{eqnarray}
where $\lim_{n\to \infty}\xi_n/n=H$.
Here the dressed energy is defined as $\xi_n(\lambda):=T\ln\eta _n(\lambda)$, with 
$\xi_n^{+}(\lambda)$ ($\xi_n^{-}(\lambda)$) denoting the dressed energy for 
$\xi _n(\lambda) \ge 0$ ($\xi _n(\lambda)<0$).  
The free energy is now
\begin{eqnarray}
f_{XXX}(T,H)&=&-\frac{1}{2} H +\sum_{n=1}^{\infty}\int_{-\infty}^{\infty}
a_n(\lambda)\xi^-_n(\lambda)d\lambda\nonumber\\
&=&J\ln 2-\int_{-\infty}^{\infty}d\lambda s(\lambda)\xi_1^+(\lambda).
\end{eqnarray}
For the ferromagnetic case $J>0$, the solution of the TBA
(\ref{TBA-XXX-T0}) is given by \cite{Takahashi}
\begin{equation}
\xi_n(\lambda)=\xi_n^+(\lambda)=2\pi Ja_n(\lambda )+  Hn \label{eta-T0} 
\end{equation}
where $n=1,2,\ldots, \infty$. 
Here we see that for $T=0$, $f_{XXX}=-H/2$. 
Therefore it follows that the fully-polarized state forms a ferromagnetic ground state. 
In this way the TBA gives rise to a direct proof of the existence of the ferromagnetic ground state.

In order to understand spin effects in the spinor Bose gas, we first discuss the low temperature behaviour of the
spinless Bose gas.  
It was for this model that Yang and Yang introduced the TBA formalism, with result 
\begin{equation}
\epsilon (k)=
\epsilon^0(k) -\mu-
\frac{T}{2\pi}\int_{\infty}^{\infty}dk' \frac{2c}{c^2+(k-k')^2}\ln(1+{\mathrm e}^{-{\epsilon(k')}/{T}})
 \label{TBA-S0}
\end{equation}
which is the special case of the TBA equation (\ref{TBA}) for the spinor Bose gas. 
At zero temperature, one can obtain physical quantities, such as the ground state energy
per unit length $E_0$, chemical potential $\mu$, pressure $P_0$  and  the cut-off momentum $Q$.
In the strong coupling limit the results for these quantities are 
\begin{eqnarray}
&&E_0\approx \frac{1}{3}n^3\pi^2\left(1-\frac{4}{\gamma}\right),\,\,\,\,
\mu_0 \approx n^2\pi^2 \left(1-\frac{16}{3\gamma}\right),\nonumber\\
&&P_0\approx \frac{2}{3}n^3\pi^2\left(1-\frac{6}{\gamma}\right),\,\,\,\,
Q \approx n\pi \left(1-\frac{2}{\gamma}\right).\label{TBA-r}
\end{eqnarray}  
The macroscopic velocity is
\begin{equation}
v_c=\sqrt{2\frac{\partial  P_0}{\partial n}}\approx \frac{\hbar\pi n}{m}\pi\left(1-\frac{4 }{\gamma}\right).
\label{vc}
\end{equation}

The low temperature thermodynamics can also be calculated directly from 
the pressure (\ref{Pressure-3}). 
For the spinless case, the function $A(T,0)\approx \mu(1+\frac{2P(T,0)}{c\mu})$. 
Substituting $A(T,0)$ into equation (\ref{Pressure-3}) and using the relation 
$\partial P(T,0)/\partial \mu =n$ gives the chemical potential
\begin{eqnarray}
\mu& \approx&
\mu_0\left[1+\frac{\pi^2}{12}\left(1-\frac{16}{3\gamma}\right)\left(\frac{K_BT}{\mu_0}\right)^2
+\frac{\pi^4}{36}\left(1-\frac{32}{5\gamma}\right)\left(\frac{K_BT}{\mu_0}\right)^4 \right].
\label{mu-0}
\end{eqnarray}
Here $\mu_0 \approx n^2\pi^2 \left(1-\frac{16}{3\gamma}\right)$
coincides with the result given in equation (\ref{TBA-r}). 
In terms of the degenerate temperature $\tau=K_BT/T_d$ the
free energy per unit length follows from relation (\ref{Free-E}) as
\begin{eqnarray}
F(\tau) \approx
E_0\left[1-\frac{\tau^2}{4\pi^2}\left(1+\frac{8}{\gamma}\right)-
\frac{\tau^4}{60\pi^4}\left(1+\frac{16}{\gamma}\right)\right]. 
\label{E-0}
\end{eqnarray}
The ground state energy $E_0$ agrees with the result given in (\ref{TBA-r}). 
The expression (\ref{Pressure-3}) indeed provides a simple way to derive the thermodynamics. 
In addition the results (\ref{mu-0}) and (\ref{E-0}) 
obtained for the spinless Bose gas via the TBA are in good agreement
with results derived from generalized exclusion statistics \cite{BG}. 
The results (\ref{mu-0})  and (\ref{E-0})
characterize the low temperature behaviour of the 1D spinless Bose gas
induced by the dynamical interaction in strong coupling regime. 
The specific heat $c_v$ and the entropy $S$ follow from the free energy (\ref{E-0}) as 
\begin{eqnarray}
c_v &=& -\frac{TL\, \partial^2 F(T,0)}{\partial T^2} \approx
\frac{N K_B \, \tau}{6(1-\frac{4}{\gamma})}+\frac{N K_B \, \tau^3}{15\pi^2(1-\frac{12}{\gamma})}
\label{Cv-S0}\\
S &=& -\frac{TL\, \partial F(T,0)}{\partial T} 
\approx \frac{N K_B \, \tau}{6(1-\frac{4}{\gamma})}+\frac{N K_B \, \tau^3}{45\pi^2(1-\frac{12}{\gamma})}
\label{S-S0}
\end{eqnarray}
which coincide with the results given in Refs.~\cite{BG,Bortz}. 
Here the degenerate temperature $\tau=K_BT/T_d$, with $T_d=\frac{\hbar^2 n^2}{2m}$.

The total energy per unit length follows from the relation $E(T,0)=F(T,0)+ST$, with result
\begin{eqnarray}
E(\tau)\approx
E_0\left[1+\frac{\tau^2}{4\pi^2}\left(1+\frac{8}{\gamma}\right)+
\frac{\tau^4}{20\pi^4}\left(1+\frac{16}{\gamma}\right)\right].
\label{E-0-2}
\end{eqnarray}
The ground state energy $E_0$ is as given in (\ref{TBA-r}).
{}From (\ref{E-0}) we see that 
\begin{equation}
F(T)=F(0)-\frac{\pi C(K_BT)^2}{6\hbar v_c} +O(T^2). \label{F-T}
\end{equation}
as expected from conformal field theory arguments for a critical system, i.e., 
for a system with massless excitations \cite{Affleck}
Here the central charge $C=1$ and $v_c$ is given by (\ref{vc}).
Similarly the finite-size corrections \cite{BCN} are given by
\begin{equation}
E(L,N)-Le_{\infty}=-\frac{\hbar\pi Cv_c}{6L}+O(1/L^2).
\end{equation}
Here $E(L,N)$ is the finite size ground state energy and $e_{\infty}$ is the energy
per unit length in the thermodynamic limit.

Furthermore, at low temperatures the strongly interacting spinless 1D Bose gas can be viewed
as a system of ideal particles obeying nonmutual generalized exclusion statistics (GES) with 
statistics parameter  \cite{BG,BGO} $\alpha \approx 1-2/\gamma$. 
These particles obey GES interpolating between bosons and fermions \cite{Haldane,Wu,Isakov}. 
For 1D interacting many-body systems the pairwise dynamical interaction between identical particles
is inextricably related to their statistical interaction through scattering.  
GES is thus the result of collective behaviour exhibited in 1D quantum many body systems. 
{}From the GES approach, the free energy and the total energy per unit length
are given by \cite{BG}
\begin{eqnarray}
F(\tau) &\approx& 
E_0\left[1-\frac{\tau^2}{4\pi^2}\left(1+\frac{2}{\gamma}\right)
+\frac{3{\bf \xi}(3)\tau^3}{2\gamma
  \pi^6}\left(1+\frac{4}{\gamma}\right)-\frac{\tau^4}{60\pi^4}\left(1+\frac{4}{\gamma}\right)\right]
\label{F-ges}\\
E(\tau) &\approx& 
E_0\left[1+\frac{\tau^2}{4\pi^2}\left(1+\frac{2}{\gamma}\right)
-\frac{3{\bf \xi}(3)\tau^3}{\gamma \pi^6}\left(1+\frac{4}{\gamma}\right) 
+\frac{\tau^4}{20\pi^4}\left(1+\frac{4}{\gamma}\right)\right].
\label{E-ges}
\end{eqnarray}
Here $\zeta(3)=\sum_{n=1}^{\infty}1/n^3$.

\section{Low temperature ferromagnetic behaviour}
\label{sec:LT}

The low temperature behaviour of the spinor Bose gas, triggered by the ferromagnetic
spin-spin interaction, is intimately related to the thermodynamic 
behaviour of the ferromagnetic Heisenberg chain, which has been extensively
studied via various methods, e.g., numerics \cite{Baker,Fisher}, spin wave
theory \cite{Takahashi2} and the TBA approach with extrapolation \cite{Takahashi4}. 
Although there has been a wide range of interest in the ferromagnetic Heisenberg chain, 
realization of ferromagnet chains are relatively rare \cite{FH-review}.
Most recent interest in the ferromagnetic Heisenberg chain has been from the perspective 
of string theory \cite{review,hagedorn}.
Obtaining exact analytical results for the thermodynamics of this model 
still provides a number of open challenges.
For one, it is extremely hard to solve the infinitely many equations  (\ref{TBA-XXX}) 
involved in the TBA.  
Nevertheless, Takahashi and his coworkers \cite{Takahashi2,Takahashi4} have given 
some results for the free energy and susceptibility which are generally accepted. 
Schlottmann \cite{Schlot2} has also predicted the leading order
of the specific heat and zero-field susceptibility via analysis of the
string solutions to the TBA equations (\ref{TBA-XXX}). 
With the help of these known results for the ferromagnetic Heisenberg chain, we show here
that,  in the strong coupling regime,
 the ferromagnetic state induced by the internal spin-spin interaction significantly affects 
the low temperature behaviour of the two-component spinor Bose gas.

\subsection{Paramagnet: $\gamma \to \infty$}

We first consider the extreme case $\gamma \to \infty$, or say $\gamma \gg 1/K_BT$. 
In this case the driving term in the TBA equations (\ref{TBA-XXX}) vanishes as $\gamma \to \infty$.  
Thus the string solutions are given by \cite{Takahashi}
\begin{equation}
\eta_n(\lambda) \approx \left[ \frac{\sinh(\frac{(n+1)H}{2T})}{\sinh(\frac{H}{2T})} \right]^2-1\label{eta}
\end{equation} 
which are known as free spin solutions. 
In this case, the particles with down-spins are unable to exchange their positions with the particles
with up-spins. 
The spins are thus frozen locally and the spin-spin exchange interaction vanishes. 
In this case, the statistical interaction is completely suppressed due to the strong repulsion. 
In this sense, for $\gamma \to \infty$ both the spinor Bose gas and the Fermi gas behave like a free
spin paramagnet.  
In this case, the spins are very sensitive to external magnetic fields, with a 
small field able to polarize all atoms.  
{}From equation (\ref{TBA-HB}) we obtain
\begin{eqnarray}
\epsilon(k) =\frac{\hbar^2 k^2}{2m}-\mu -K_BT\ln \left(2\cosh \frac{H}{2K_BT}\right).
\end{eqnarray}
Under the condition $\gamma \gg 1/K_BT$ and $K_BT< T_d$, the free energy (\ref{Free-E}) gives 
\begin{eqnarray}
F(\tau,h) &\approx &
\frac{1}{3}\frac{\hbar^2}{2m}\pi^2n^3\left[1-\frac{\tau^2}{4\pi^2}-\frac{\tau^4}{60\pi^4}
 -\frac{3\tau}{\pi^2}\ln \left(2\cosh \frac{h}{2\tau}\right)\right]
 \label{F-FS}
\end{eqnarray}
where we have set $h={H}/{T_d}$.

\begin{center}
\begin{figure}
\includegraphics[width=0.95\linewidth]{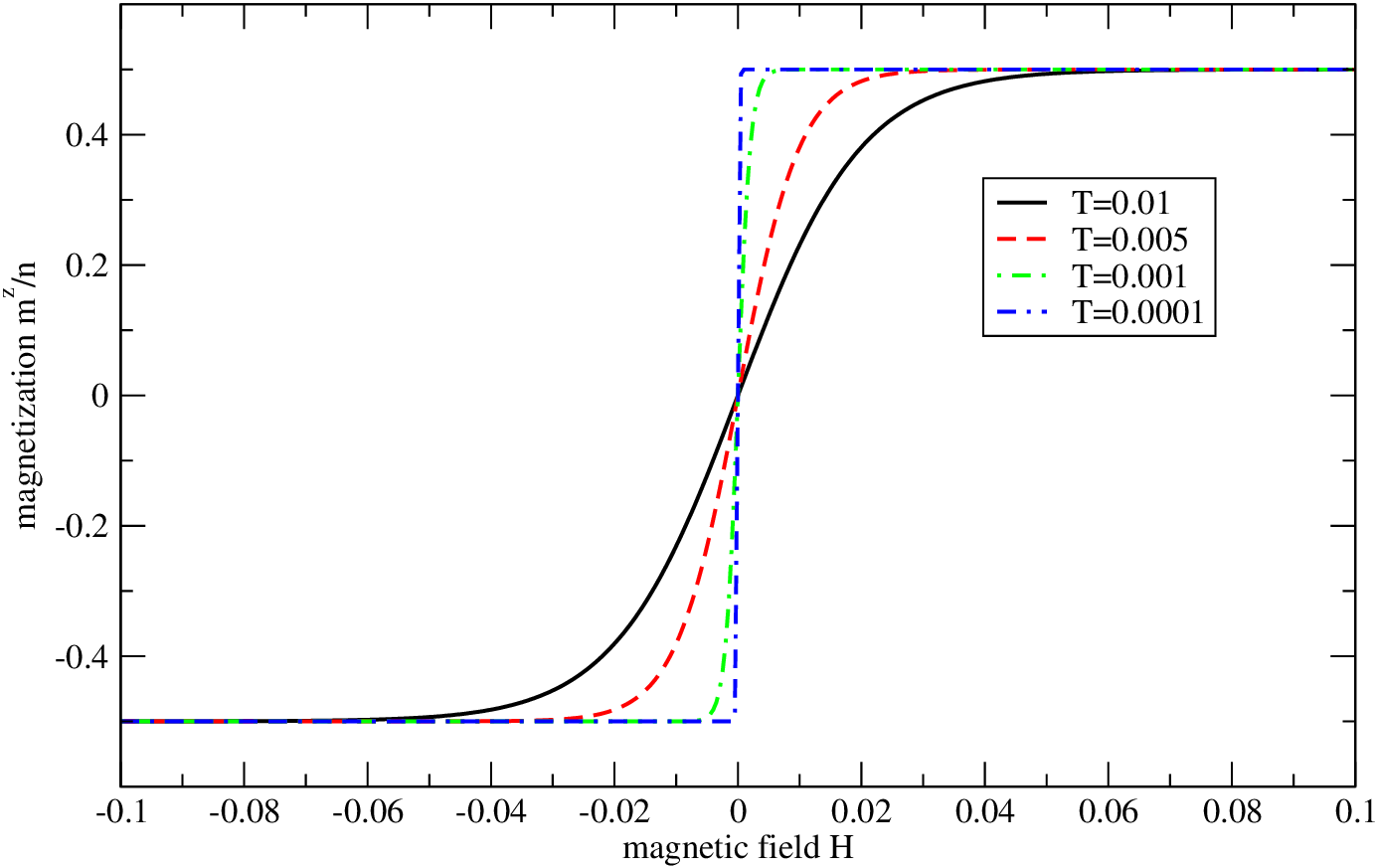}
\caption{(color online) Magnetization (normalized by the linear density $n$) vs magnetic
  field $H$ (with Bohr magneton $\mu_B=1$) for different temperatures
  (in units of $K_B$). In the strong coupling limit $\gamma \to
  \infty$ two-component spinor bosons exhibit free spin behaviour at
  low temperatures.}
\label{fig:mzS0}
\end{figure}
\end{center}

It follows that in the strong coupling limit $\gamma \to \infty$ the magnetic properties of the
two-component spinor Bose gas are those of $su(2)$ free spins with a
divergent susceptibility $\chi \approx \frac{n}{4K_BT}\left(1-\tanh^2(H/(2K_BT))\right)$ at low temperatures. 
The magnetization is $m^z=\frac{1}{2} n \tanh (H/(2K_BT))$. 
Figure \ref{fig:mzS0} shows the free spin behaviour in the magnetization as a function of the magnetic field at different temperatures. 

\begin{center}
\begin{figure}
\includegraphics[width=0.95\linewidth]{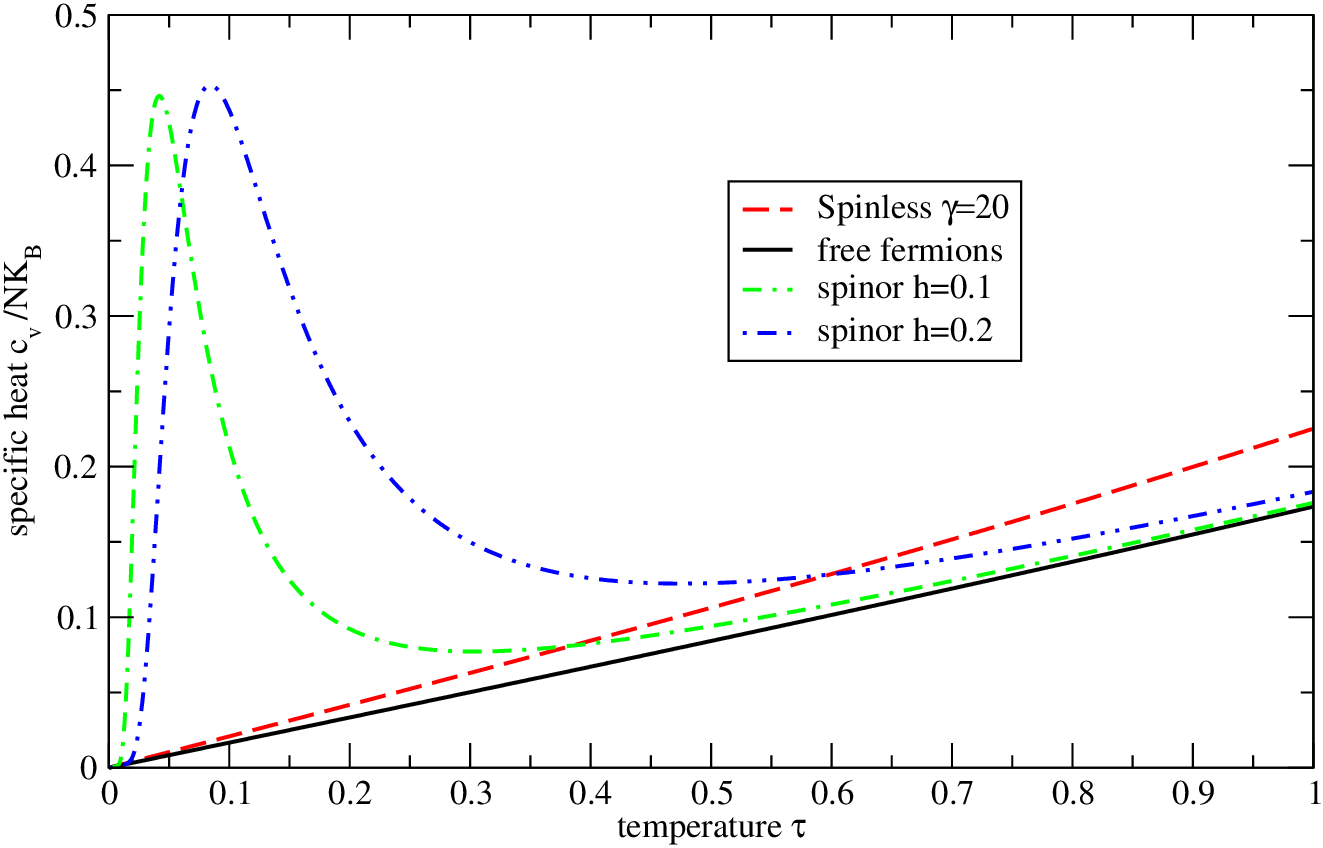}
\caption{(color online) Specific heat normalized by $NK_B$ vs degenerate temperature $\tau=K_BT/T_d$.  
The straight dashed line is the specific heat (\ref{Cv-S0}) for the spinless
  Bose gas with $\gamma=20$.  The solid line is the
  specific heat for the free fermions. The curves show the specific heat
  (\ref{Cv-f}) of the spinor Bose gas with $\gamma \to \infty$ for finite field term $h$.}
\label{fig:cvS0}
\end{figure}
\end{center}

The specific heat and entropy follow from equation (\ref{F-FS})  as
\begin{eqnarray}
c_v &\approx&
\frac{1}{6}{N K_B \tau}+\frac{N K_B\tau^3}{15\pi^2}
+\frac{N K_B h}{4\tau^2}^2\left[ 1-\tanh^2\left(\frac{h}{2\tau}\right)\right]
\label{Cv-f}\\
S &\approx&\frac{1}{6}{N K_B \tau}+\frac{N K_B \tau^3}{45\pi^2}+N K_B \ln\left[2\cosh
\left(\frac{h}{2\tau}\right)\right]-\frac{N K_B h}{2\tau}\tanh\left(\frac{h}{2\tau}\right).\label{S-f}
\end{eqnarray}
In the  absence of external field ($H=0$) the specific heat behaves like that of a free Fermi gas. 
However, the free spins make a contribution $N K_B \ln2$  to the free Fermi entropy. 
Figure \ref{fig:cvS0} shows a plot of the specific heat in the presence of magnetic field. 
We see clearly here in the strong coupling limit that the specific heat
is sensitive to the external field. 
We also note that our calculations differ from those in Refs~\cite{Li,TBA-Li} where the spin degrees of freedom
were ignored.

\subsection{Ferromagnetic: $1\ll \gamma<1/K_BT$}

For finite temperatures the solutions of the TBA equations (\ref{TBA-XXX}) vary from the 
free spin solutions (\ref{eta-T0}) and (\ref{eta}) \cite{Takahashi,Schlot2}. 
The spin-spin exchange interactions are enhanced as the interaction strength $\gamma$ decreases 
from the strong coupling limit.
In this section we explore the ferromagnetic behaviour of  the spinor Bose gas with finitely strong interaction,
i.e., in the regime $1\ll \gamma<1/K_BT$, or more precisely speaking,
$\frac{c}{2P(T,0)}<1/K_BT$. 
In this regime, the known results for the free energy of the ferromagnetic Heisenberg chain 
at low temperatures obtained by Takahashi and colleagues are applicable to the ferromagnetic state 
associated with the free energy (\ref{TBA-XXX-f}).
The result is \cite{Takahashi2,Takahashi4}
\begin{eqnarray}
f_{XXX}(T,0) &\approx & J\left[-1.042
\left(\frac{K_BT}{J}\right)^{\frac{3}{2}}+\left(\frac{K_BT}{J}\right)^2 - 0.9\left(\frac{K_BT}{J}\right)^{\frac{5}{2}}\right] 
\label{f-XXX-T}
\end{eqnarray}
with effective coupling strength $J\approx 2P(T,0)/c$.  
At low temperature, the effective coupling approaches the constant value $J\approx 4E_{\rm F}/3\gamma$ as a result of the
temperature-dependent part in the coupling only making an $O((k_BT)^3)$ contribution to the free energy 
of the spinor Bose gas.
Schlottmann \cite{Schlot2} calculated the leading terms of the
specific heat and free energy for the ferromagnetic Heisenberg chain
by approximating the infinite set of TBA equations (\ref{TBA-XXX}). 
Using Schlottmann's method we find from the TBA equations (\ref{TBA-XXX}) that the leading order 
in the free energy is proportional to $(K_BT)^{\frac{3}{2}}$. 
With the help of Takahashi's result for the ferromagnetic Heisenberg chain (\ref{f-XXX-T}), we may
calculate the thermodynamics of the spinor Bose gas within finitely strong interaction, which we now do.

Substituting $A(T,0)=\mu+{2P(T,0)}/{c}-f_{XXX}(T,0)$ into equation (\ref{Pressure-3}) we have 
\begin{eqnarray}
P(T,0) &\approx &
\frac{1}{\sqrt{\frac{\pi^2\hbar^2}{2m}}}\frac{2\tilde{\mu}^{\frac{3}{2}}}{3}\left[
  1+\frac{2}{c}\frac{\sqrt{\tilde{\mu} }}{\sqrt{\frac{\pi^2\hbar^2}{2m}}}
+\frac{\pi^2}{8}\left(\frac{K_BT}{\tilde{\mu}}\right)^2
+\frac{\pi^2}{6}\left(\frac{K_BT}{\tilde{\mu} }\right)^2
\frac{\sqrt{\tilde{\mu }}}{c\sqrt{\frac{\pi^2\hbar^2}{2m}}}\right]
\end{eqnarray}
where $\tilde{\mu} =\mu -f_{XXX}(T,0)$. 
For the regime $\frac{c}{2P(T,0)}<1/K_BT$ calculation of the chemical potential
via the relation $\partial P(T,0)/\partial \mu =n$ gives 
\begin{eqnarray}
\mu &\approx& \mu_0\left[1+\mu_1\left( \frac{\gamma
  K_BT}{\mu_0}\right)^{\frac{3}{2}}+\mu_2\left( \frac{\gamma
  K_BT}{\mu_0}\right)^{2}+\mu_3\left( \frac{\gamma K_BT}{\mu_0}\right)^{\frac{5}{2}}+ 
  \frac{\pi^2}{12}\left(1-\frac{16}{3\gamma}\right)\left( \frac{K_BT}{\mu_0}\right)^{2}\right]
\end{eqnarray}
where 
\begin{eqnarray}
\mu_1= \frac{1.042\times \sqrt{3}}{4\gamma}\left(1-\frac{17}{3\gamma}\right),\quad
\mu_2 =-\frac{3}{2\gamma}\left(1-\frac{7}{3\gamma}\right), \quad 
\mu_3 =\frac{0.9\times 21\sqrt{3}}{16\gamma}\left(1-\frac{11}{7\gamma}\right).\nonumber
\end{eqnarray}

Some algebra gives the free energy $F(\tau,0)$ and total energy $ E(\tau,0)$ 
per unit length
\begin{eqnarray}
F(\tau,0) &\approx&  E_0\left[1-\frac{1.042\times
3\sqrt{3}}{2\gamma\pi^3}\left(
1+\frac{7}{\gamma}\right)(\gamma\tau)^{\frac{3}{2}} 
 +
\frac{9}{4\gamma\pi^4}\left(1+\frac{10}{\gamma}\right)(\gamma\tau)^{2}-\frac{0.9\times 9\sqrt{3}}{8\gamma\pi^5}\left(
1+\frac{13}{\gamma}\right)(\gamma\tau)^{\frac{5}{2}} \right.\nonumber\\
& & \qquad \left.
 -\frac{\tau^2}{4\pi^2}\left(1+\frac{8}{\gamma}\right)\right]  \label{F-H}\\
 E(\tau,0) &\approx& E_0\left[ 1+\frac{1.042\times
3\sqrt{3}}{4\gamma\pi^3}\left(
1+\frac{7}{\gamma}\right)(\gamma\tau)^{\frac{3}{2}}  -
\frac{9}{4\gamma\pi^4}\left(1+\frac{10}{\gamma}\right)(\gamma\tau)^{2}
+\frac{0.9\times 27\sqrt{3}}{16\gamma\pi^5}\left(
1+\frac{13}{\gamma}\right)(\gamma\tau)^{\frac{5}{2}}
\right.\nonumber\\
& & \qquad \left. +\frac{\tau^2}{4\pi^2}\left(1+\frac{8}{\gamma}\right)\right]. \label{E-H}
\end{eqnarray}
We see explicitly that the ferromagnetic behaviour of the spin exchange
interaction dominates the thermodynamics of the strongly interacting two-component Bose gas.

\begin{center}
\begin{figure}
\includegraphics[width=0.8\linewidth]{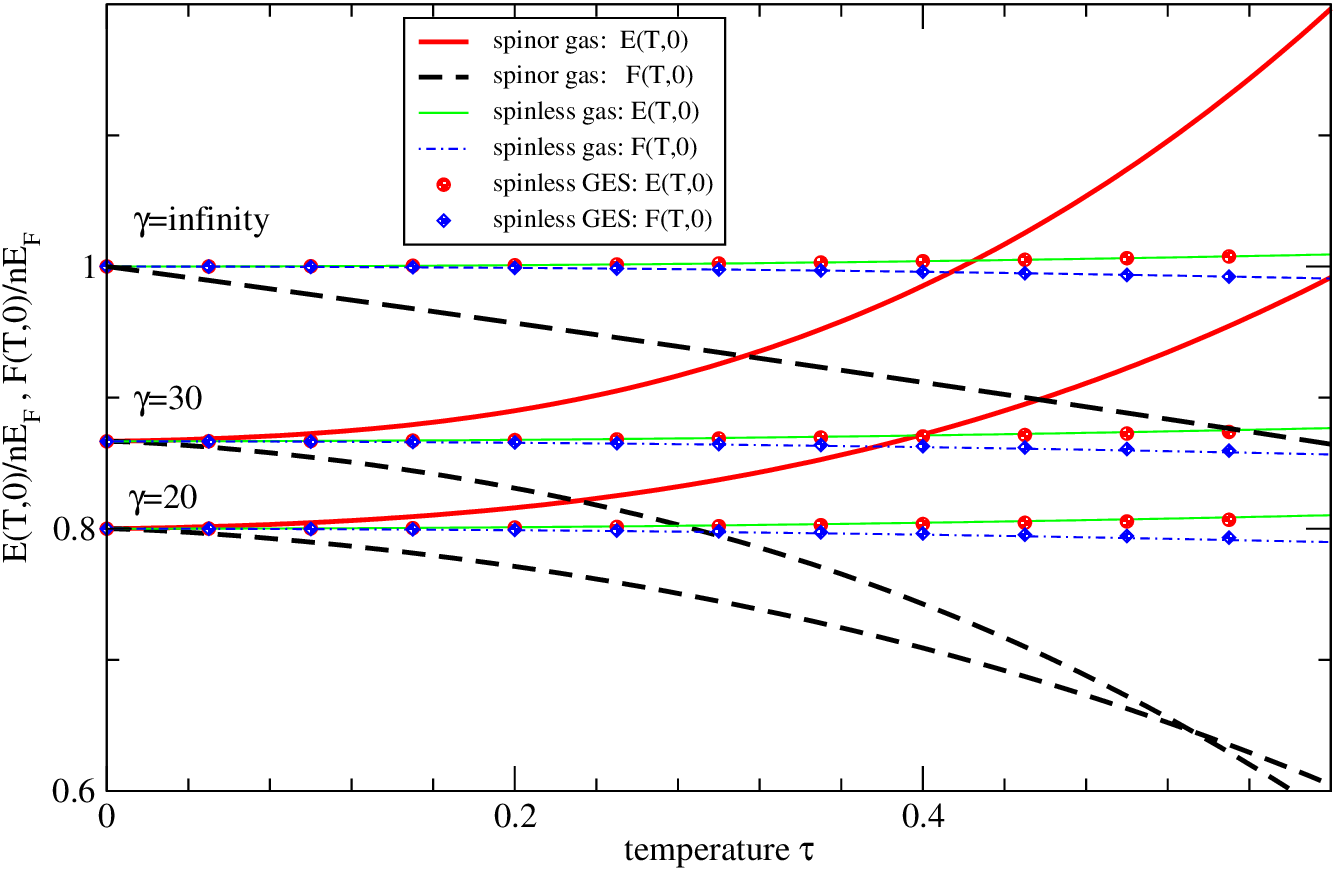}
\caption{(color online) 
Free energy $F(T,0)$  and total energy $E(T,0)$ per unit length (normalized by  
$nE_{\rm F}=\frac{\hbar^2}{2m}\pi^2n^3$) as a function of the degenerate temperature $\tau=K_BT/T_d$ 
for different coupling strengths for both the two-component spinor and spinless 1D Bose gases.
The thick solid and dashed  lines are the energies (\ref{F-H}) and (\ref{E-H}) of the spinor Bose gas. 
The flatter thin solid and dashed lines are the results (\ref{E-0}) and (\ref{E-0-2}) for the spinless Bose gas.
For further comparison the symbols show the energies (\ref{F-ges}) and (\ref{E-ges}) derived 
for the spinless Bose gas from an ideal gas of particles obeying nonmutual GES \cite{BG}. 
In general the curves reveal the universal characteristics of the energies at low temperatures.
}
\label{fig:E-F-all}
\end{figure}
\end{center}

It is of interest to note that the leading temperature-dependent
term in the energy expressions is $O(T^{3/2})$ for the spinor Bose gas 
compared to $O(T^2)$ for the spinless Bose gas (\ref{E-0}). 
In the strong interaction regime $1\ll \gamma< \mu_0/K_BT$ this  
leads to quantitatively different low temperature behaviour compared to the spinless Bose gas.
Figure \ref{fig:E-F-all} shows the total energy and the free
energy in the strong interaction regime as a function of temperature for the spinor Bose gas, the spinless Bose gas 
and an ideal gas obeying GES. 
Significantly different characteristics of low temperature behaviour for the spinor Bose gas
and the spinless Bose gas are depicted.  
For the spinless Bose gas with strong coupling, the low temperature thermodynamics is known to 
coincide with that of ideal particles obeying nonmutual GES with statistics parameter 
$\alpha \approx 1-2/\gamma$. 
Figure \ref{fig:E-F-all} hints that the strongly interacting spinor Bose gas might also 
be equivalent to a gas of ideal particles obeying nonmutual GES.  
It is an open question as to what the statistics parameter $\alpha$ for the spinor Bose
gas might be.

\begin{center}
\begin{figure}
\includegraphics[width=0.95\linewidth]{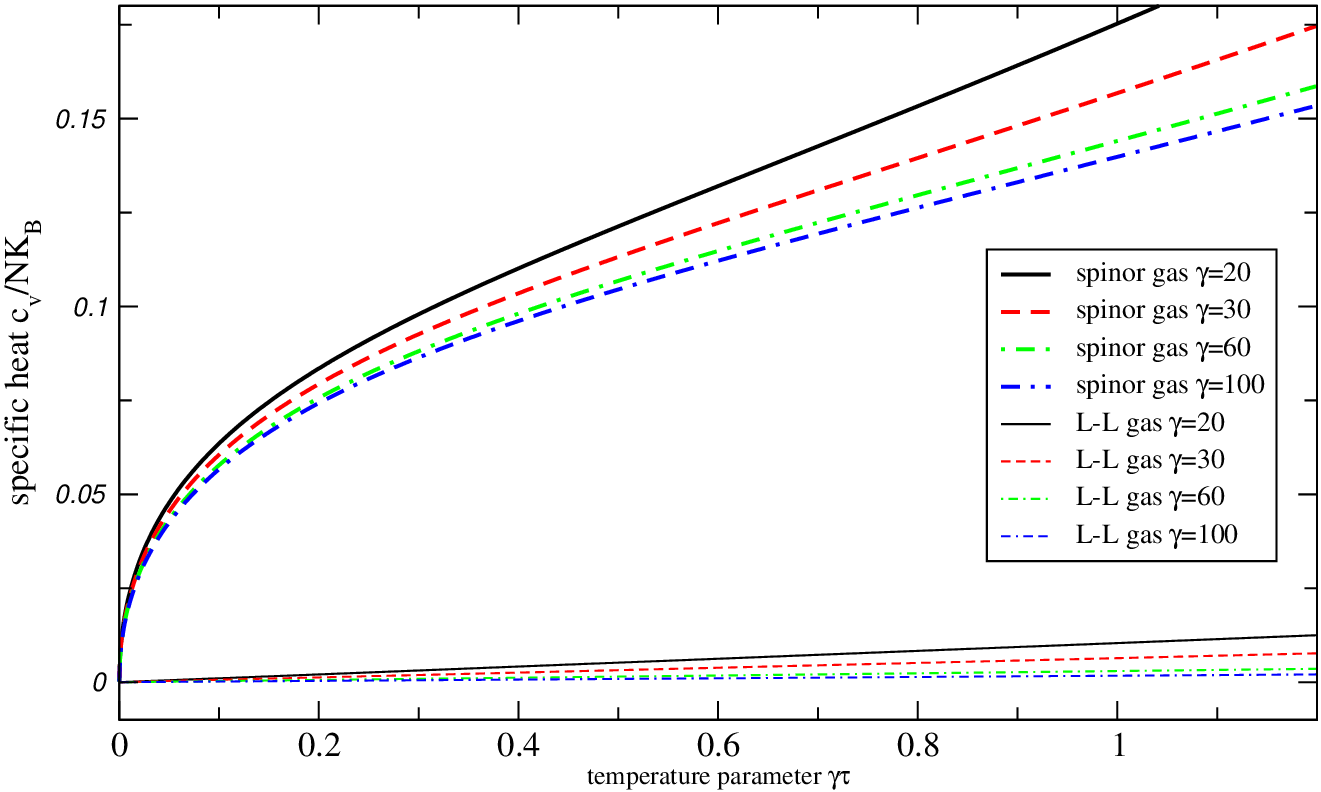}
\caption{(color online) 
Comparison between the specific heat (normalized  by $NK_B$) of the two-component spinor Bose gas and the 
spinless Bose gas as a function of the temperature parameter $\gamma\tau$ at different 
coupling strengths. 
The upper set of curves are obtained from the spinor Bose gas specific heat (\ref{cv-S}).
The lower set of curves follow from the spinless Bose gas result  (\ref{Cv-f}).
In general the curves highlight the distinction between the temperature-dependent behaviour 
of the spinor and spinless Bose gases.
 }
\label{fig:cv-all}
\end{figure}
\end{center}

To conclude this section the specific heat and entropy  are  given by
\begin{eqnarray}
\frac{c_v}{N K_B} & \approx & \frac{1.042\times
3\sqrt{3}(\gamma \tau)^{\frac{1}{2}}}{8(1-\frac{3}{\gamma})\pi}-\frac{3(\gamma
\tau)}{2(1-\frac{6}{\gamma})\pi^2}+\frac{0.9\times
45\sqrt{3}(\gamma \tau)^{\frac{3}{2}}}{32(1-\frac{9}{\gamma})\pi^3}
+\frac{\tau}{6(1-\frac{4}{\gamma})}
\label{cv-S} \\ 
\frac{S}{N K_B} & \approx &  \frac{1.042\times
3\sqrt{3}(\gamma \tau)^{\frac{1}{2}}}{4(1-\frac{3}{\gamma})\pi}-\frac{3(\gamma
\tau)}{2(1-\frac{6}{\gamma})\pi^2}+\frac{0.9\times
15\sqrt{3}(\gamma \tau)^{\frac{3}{2}}}{16(1-\frac{9}{\gamma})\pi^3} +\frac{\tau}{6(1-\frac{4}{\gamma})}
\end{eqnarray} 
which differ significantly from the corresponding spinless Bose gas results 
(\ref{Cv-S0}) and (\ref{S-S0}) and also from free spin case (\ref{Cv-f}) and (\ref{S-f}). 
Figure \ref{fig:cv-all} shows the specific heat as a function of the temperature. 
The specific heat exponent $c_v\simeq T^{-a} $
indicates that $a=-0.5$ for spinor Bose gas for the regime
$1\ll \gamma<\mu_0/K_BT$ whereas $a=-1$ for the Lieb-Lininger gas. For
$\gamma\gg 1/K_BT$, the spinor Bose gas and the Lieb-Lininger gas both
have $a=-1$ for the absence of the external field.

\subsection{Susceptibility: $1\ll\gamma< 1/K_BT$}

Now we consider the effect of a small external field ($H\ll T$) within the regime $1\ll\gamma< 1/K_BT$. 
Here we adapt the known free energy result for the
ferromagnetic chain in the presence of an external field \cite{Takahashi2,Takahashi4}, namely  
\begin{eqnarray}
f_{XXX}(T,H)\approx f_{XXX}(T,0) -\frac{H^2 J}{8(K_BT)^2}
\left[ \frac{1}{6}+ 0.5826
\left(\frac{K_BT}{J}\right)^{\frac{1}{2}}+0.678 \left(\frac{K_BT}{J}\right)\right]
\label{f-XXX-T-H}
\end{eqnarray}
with $f_{XXX}(T,0)$ as given in (\ref{f-XXX-T}).
%
%
Repeating the procedure of the previous section with the free energy (\ref{f-XXX-T-H}) gives the result 
\begin{eqnarray}
F(\tau,h) &\approx&  F(\tau,0)
-\frac{E_0h^2}{\tau^2}\left[ \frac{1}{12\gamma} \left(1-\frac{2}{\gamma}\right)+\frac{0.5826\times
  \sqrt{3}}{4\gamma \pi} \left(1-\frac{1}{\gamma}\right) \left(
  \gamma\tau\right)^{\frac{1}{2}}+\frac{0.678\times
  3}{8\gamma\pi^2} \left(1+\frac{4}{\gamma}\right) \left(\gamma\tau\right)\right] 
  \label{F-H-2}
\end{eqnarray}
with $F(\tau,0)$ as given in (\ref{F-H}).

\begin{center}
\begin{figure}
\includegraphics[width=0.95\linewidth]{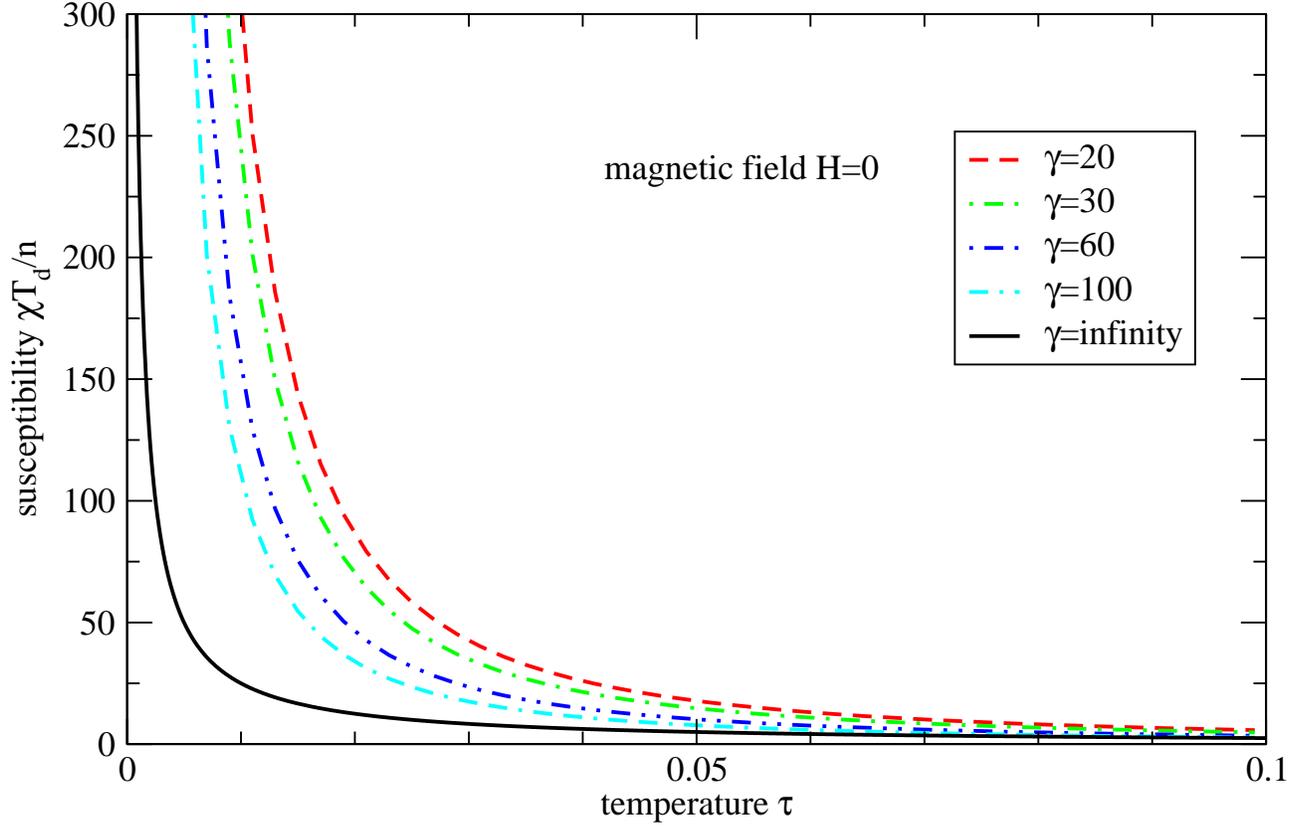}
\caption{(color online) 
Susceptibility (normalized by  $n/T_{\rm d}$) as a function of the degenerate temperature $\tau$ for different
values of interaction strength $\gamma$. 
In the strong coupling limit $\gamma \to \infty$ the two-component spinor bosons behave like a
paramagnet with susceptibility $\chi \sim T^{-1} $. 
For finitely strong interaction, the ferromagnetic susceptibility behaves as 
$\chi \sim T^{-2}$.
 }
\label{fig:chi}
\end{figure}
\end{center}

For small magnetic field ($h<\tau$) the susceptibility per unit length 
\begin{eqnarray}
\chi = -\frac{\partial^2 F(T,H)}{\partial H^2} 
\approx  \frac{ n}{T_d}\left[ \frac{\pi^2}{18\gamma\tau^2} \left(1-\frac{6}{\gamma}\right)+\frac{0.5826\times
  \sqrt{3}\pi}{6\sqrt{\gamma}\tau^{\frac{3}{2}}} \left(1-\frac{3}{\gamma}\right)+\frac{0.678}{4\tau}  \right]
\end{eqnarray}
is indeed greater than that of free spins, for which $\chi\approx n/(4K_BT)$. 
%
%
This result is also consistent with Eisenberg and Lieb's general argument for polarized spinor bosons \cite{Eisenberg-Lieb}. 
Figure \ref{fig:chi} shows the zero-field susceptibility as a function of temperature for different values of the 
interaction strength. 
The susceptibility diverges as $\tau \to 0$. 
The susceptibility decreases with increasing interaction towards the 
free spin paramagnetic susceptibility is the lowest curve (solid line). 
%
%
%
The susceptibility exponent defined by $\chi \sim T^{-b}$ is $b=2$ for the regime $\gamma< \mu_0/K_BT$ with 
$b=1$ for the paramagnet.

\section{Local pair correlation}
\label{sec:pair}

The local pair correlation function for the 1D Bose gas has been determined experimentally 
in a gas $^{87}$Rb atoms as a function of the interaction strength by measuring photoassociation rates  \cite{Weiss}.  
In general local two-particle correlations can be used to study phase coherence behaviour and 
classify various finite temperature regimes in 1D interacting quantum gases \cite{Shlyapnikov,Cazalilla}. 
In the grand canonical description, two-particle pair correlation are 
given in terms of the  field operator $\Psi$  and the free energy $f(\gamma,T)$ 
by \cite{Korepin,Shlyapnikov}
\begin{equation}
g^{(2)}(0):=\langle \Psi^{\dagger}\Psi^{\dagger}\Psi\Psi \rangle
=\frac{2m}{\hbar^2n^2}\left(\frac{\partial
  f(\gamma,T)}{\partial \gamma }\right)_{n,t}.
\end{equation} 
At zero temperature the local pair correlation is $g^{(2)}\approx 1$ for the weakly interacting Bose gas.  
On the other hand, $g^{(2)} \rightarrow 0$ as $\gamma$ increases into the Tonks-Girardeau regime, 
indicative of free fermionic behaviour. 
In this regime, the long range behaviour is characterized by the one-body correlation
function $g^{(1)}(x) =\langle \Psi^{\dagger}(x)\Psi(0)\rangle \propto 1/\sqrt{x}$, 
corresponding to the momentum distribution $n(p)\propto 1/\sqrt{p}$. 
Here $p$ is the momentum. 
More generally, in terms of the Luttinger parameter $K$ \cite{CAZA1}, 
$g^{(1)}(x) \propto 1/x^{1/2K}$ and $n(p) \propto 1/p^{(1-1/2K)}$. 
In the weak coupling limit $K\approx \pi/\sqrt{\gamma}$, which leads to a
power-law decay in the one point correlation.

The pair correlation function for the spinless Bose gas 
\begin{equation}
g^{(2)}(0)\approx
\frac{4\pi^2}{3\gamma^2}\left(1+\frac{\tau^2}{4\pi^2}
+\frac{\tau^4}{20\pi^4}\right)
\end{equation}
follows from the TBA result (\ref{E-0}), which coincides with the result given in Refs.~\cite{Shlyapnikov,Cazalilla}.
It is evident that the dynamical interaction dramatically reduces pair correlation
due to decoherence between individual wave functions of colliding particles.
On the other hand, increasing temperature slowly enhances local pair correlation.  
At temperatures $\tau \ll 1$ ($T \ll T_d$), the local pair correlation approaches 
free Fermi behaviour as $\gamma \to \infty$, as was 
quantitatively demonstrated in the experimental observations of the pair correlation function 
for a gas of interacting $^{87}$Rb atoms confined to 1D \cite{Weiss}.

\begin{center}
\begin{figure}
\includegraphics[width=0.95\linewidth]{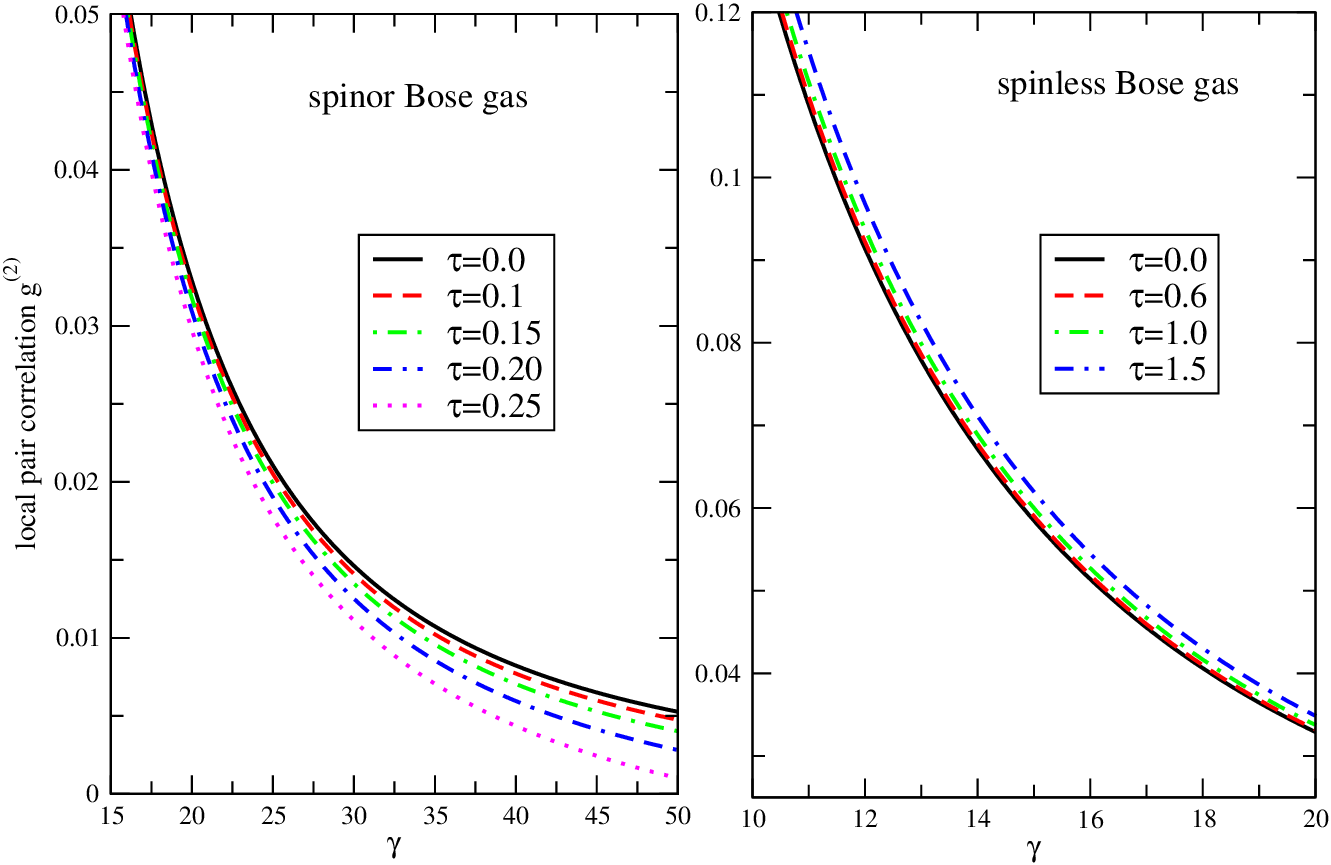}
\caption{(color online) 
Local pair correlation  $g^{(2)}$ as a function of interaction strength $\gamma$ at 
different temperatures.
For the spinless Bose gas, the pair correlation function increases with increasing temperature. 
In contrast, for the two-component spinor Bose gas the pair correlation function decreases with increasing 
temperature as a result of the temperature induced spin-spin ferromagnetic exchange interaction.
}
\label{fig:g2}
\end{figure}
\end{center}

For the two-component spinor Bose gas considered here, the ferromagnetic spin-spin exchange
interaction results in a different temperature-dependent pair correlation function.
For this model, the local pair correlation function 
\begin{eqnarray}
g^{(2)}(0)&\approx &
\frac{4\pi^2}{3\gamma^2}\left[1-\frac{1.042\times 3\sqrt{3}(\gamma\tau)^{\frac{3}{2}}}{16\pi^3}\left(1-\frac{3}{\gamma}\right)
+\frac{9(\gamma\tau)^{2}}{16\pi^4} -\frac{0.9\times
  27\sqrt{3}(\gamma\tau)^{\frac{5}{2}}}{64\pi^5}\left(1+\frac{3}{\gamma} \right)+\frac{\tau^2}{4\pi^2} \right]
\end{eqnarray}
follows from equation (\ref{F-H}) in the regime $\mu_0/K_BT > \gamma \gg 1$.
We see for the spinor Bose gas the local pair correlation again quickly decays with respect to
the dynamical interaction $\gamma$. 
Figure (\ref{fig:g2}) shows the local pair correlation for the spinless and the spinor
Bose gases as a function of the interaction strength at different temperatures.  
In contrast to the spinless Bose gas, where the temperature enhances
local pair correlation due to thermal fluctuations, for the spinor Bose gas 
the local pair correlation decreases with increasing temperature due to the 
ferromagnetic spin-spin exchange interaction.

\section{Conclusion}
\label{sec:concl}

In this paper we have studied the thermodynamics of the integrable 1D
two-component Bose gas via the thermodynamic Bethe ansatz.
Analytic low temperature results were obtained for the free energy, total energy, specific heat, 
entropy, pressure, susceptibility and pair correlation function in the strongly interacting regimes 
$\gamma\gg 1/K_BT$ and $1 \ll \gamma< \mu_0/K_BT$. 
Where appropriate, comparison was made with corresponding thermodynamic properties of the integrable 
1D spinless Bose gas.
Our key finding is that the temperature induced ferromagnetic spin-spin 
exchange interaction triggers a number of novel quantum effects 
in the thermodynamic properties of the spinor Bose gas at low
temperatures. 
In the regime $1 \ll \gamma< \mu_0/K_BT$
the specific heat exponent for the spinor Bose gas following from 
$c_v\sim T^{1/2}$ is different to that of the spinless  Bose gas for which $c_v\sim T$. 
In this regime, the susceptibility exponent is given by $\chi \sim T^{-2}$ which
exceeds that of free spin paramagnet for which $\chi \sim T^{-1}$. 
In contrast to the spinless Bose gas, where the pair correlation function increases with increasing temperature, 
the two-component spinor Bose gas pair correlation function decreases with increasing 
temperature as a result of the temperature induced spin-spin ferromagnetic exchange interaction.

In general these exact results should be relevant to understanding ferromagnetic behaviour and spin effects 
in two-component spinor Bose gases of cold atoms, for which the interaction strength can in principle be tuned.
However, the introduction of precise thermometry into these systems to measure universal temperature 
dependent effects provides a number of challenges. 
However, this is a worthwhile goal.
As we have seen, for strong coupling the characteristics of the thermodynamics of the spinor Bose gas at low
temperatures are described by an effective ferromagnetic Heisenberg spin chain. 
There is  a remarkable three-way correspondence  \cite{review,hagedorn} between the 
ferromagnetic Heisenberg chain, 
the limit of weakly coupled planar ${\cal N} = 4$ supersymmetric Yang-Mills theory and the limit of 
free strings on ${\rm AdS}_5 \times {\rm S}^5$.
This triality between guage theory, string theory and the thermodynamics of the ferromagnetic Heisenberg chain
has recently been used to calculate the Hagedorn temperature of the string theory in agreement with the 
Hagedorn/deconfinement temperature calculated on the guage theory side \cite{hagedorn}.
It now appears that we can add a further connection with the thermodynamics of the 
strongly interacting two component spinor Bose gas.

\acknowledgments

This work is supported by the Discovery and Linkage International programs of the 
Australian Research Council through grants DP0342561, DP0663773 and LX0455823.
It is also in part supported by Grant-in-Aid for the Scientific
Research (B) No. 18340112 from the Ministry
of Education, Culture, Sports, Science and Technology, Japan.
The authors thank Michael Bortz, Fabian Essler, Andreas Kl\"umper, Masahiro Shiroishi 
and David Weiss for helpful discussions. 
M. T. thanks the Department of Theoretical Physics for hospitality during the initial
part of this work. 
X.-W. G. and M. T. B.  thank the Institute of Solid State Physics, University of Tokyo and 
Institut Henri Poincare -- Centre Emile Borel for hospitality and support during the final stages.

\end{document}